# Bioinformatics Knowledge Transmission (training, learning, and teaching): overview and flexible comparison of computer based training approaches.

Etienne Z. Gnimpieba, Douglas Jennewein, Luke Fuhrman, Carol M. Lushbough
Computer Science Department, University of South Dakota, 414 E. Clark St. Vermillion, SD 57069, USA,
{Etienne.gnimpieba; Doug.Jennewein; Luke.Fuhrman, Carol.Lushbough}@usd.edu

**Corresponding author:** Etienne.gnimpieba@usd.edu, +1 605 223 0383.

~o~

**Abstract:** The merger of computer science, mathematics, and life sciences has brought about the discipline known as bioinformatics. However, the transmission (e.g. training, learning, and teaching) of this knowledge becomes an important issue. Many tools have been developed to help the bioinformatics community with that transmission challenge. When selecting the best of these tools, called here BKTMS (Bioinformatics Knowledge Transmission Management Systems), there may be confusion. What makes a good BKTMS? How can we make this choice efficiently? These questions remain unanswered for many users (e.g. learner, teacher and student, trainer and trainee, administrator). This paper provides a critical review of 32 existing BKTMS and a flexible comparison. This review and evaluation will be used to gain insight into the tools, systems, and capabilities that will be added to or excluded from a new proposed model for the next generation of BKTMS, involving multidisciplinary, web semantic tools (e.g. web services, workflow) and standards like LOM, or SCORM.

## Introduction

Bioinformatics allowed creation of vast knowledge amounts based on data, tools, processes, and technology. Bioinformatics aims to use and create technologies for identification, manipulation, modification, and creation of life science data. Currently, the transmission of bioinformatics knowledge and skills is a difficult task to accomplish [1]. The growing mass of bioinformatics knowledge, data, and tools has led to a growing need for adequate knowledge transmission and collaborative tools.

The diversity of academic domains, student profiles, and skills requires the specialization and the personalization of bioinformatics training programs. Over the past ten years, higher education institutions have begun to offer courses in bioinformatics and computational biology [2]. A survey conducted by Messersmith et al. shows that about 30% of United States universities offered bioinformatics educational workshops as of 2011 [3]. Some high schools in the world have begun education in bioinformatics as well [4]. In addition, we are witnessing a growing development of tools for managing the transmission of knowledge using computer engineering, called computer-based training (CBT). Users are faced with the difficulties of choice and the good definition quality of the results. In the bioinformatics field, the complexities of the subjects (i.e. multidisciplinarity) and the diversity of training needs make a choice even more difficult. Indeed, most of the management tools for bioinformatics knowledge transmission that have been developed in recent years are related to proprietary needs or specific bioinformatics software (e.g. ExPASy, EBI, NCBI, ...). Other management tools try to adapt existing educational frameworks to facilitate standardization of their content (e.g. E-Biomics, EMBR, OpenHelix, etc.). Others tried to build new transmission strategies based on Bioinformatics knowledge networking (e.g. BTN, Biostar). However, it is difficult to select the appropriate

BKTMS from this panoply of BKTMS. An evaluation of these tools proves necessary.

For example, possessing statistics skills can qualify a person for several career profiles such as *systems biology* modeling, data analysis, or data modeling (data learning). This concept of knowledge transmission requires modularity of transmitted knowledge. In fact, to transmit the knowledge necessary for understanding complex concepts in bioinformatics and computational biology such as modeling of systems biology [5], [6], a BKTMS needs modularity to facilitate both transmitting and receiving. Standardization, modularity, reusability and flexibility prove to be important in the design criteria of BKTMS (Yusof, Mansur, & Othman, 2011; [8]. This paper presents a critique of 32 existing BKTMS and a flexible score-based comparison of these BKTMS with 8 generic Learning Management System (LMS).

1. **Bioinformatics Knowledge Transmission Management System (BKTMS) overview**

A BKTMS is a web portal that provides resources for learning, teaching, training, or helping with executing bioinformatics work/problems. It can be a learning object repository (LOR), learning management system (LMS), courses management system (CMS), virtual learning environment (VLE), computer based training (CBT) portal or a simple website. These portals intend to offer a service to different groups of users in order to facilitate an understanding of specific tools, databases, functions, exercises, bioinformatics processes, programs, skills, and career profile requirements. In many cases, this service is made available for educational purposes in order to advance knowledge and understanding in bioinformatics.

There are some existing portals. What are their flaws and benefits? Table 1 indicates a number of portals reviewed with a standard set of theoretical parameters. Some descriptive parameters and information can be found in the educational tools evaluation system literature (Landon, Henderson, & Poulin, 2006, [2]. Our selected features to describe tools are: tool identification (Portal name, Base Institution, contact/author, main Statement/Goals, URL/location, comments) and technical information (approach, login, freely downloadable accessible materials, updates clearly indicated, reviewing/ranking material option, searchable materials, trainer/contact information, information about training facilities, links to courses and events).

Using these features, Table 1 describes 32 BKTMS. With some BKTMS there was no clear approach in the design or learning strategy, but instead a presentation collection of training resources. Through the accessibility evaluation criteria, BKTMS are categorized as free, partially free (free for academic use for example), and paid systems. Free tools are preponderant. To the nature and organization of content, there is a low diversity of formats (usually slides, videos or text). The organization is usually related to owner activities. BKTMS are usually organized by topics, by subdomain, by tool (software, database), on application or case study (DNA sequence analysis). This organization diversity attests to the complexity of the bioinformatics domain, but reveals that standardization of the learning context should be evaluated for efficient bioinformatics knowledge transfer.

Many of the tools and portal aspects mentioned previously are required for a competent bioinformatics training portal; however, a few new integrations would be useful as well. Many portals include profiles and sign-in capabilities, but it seems that none use that function as effectively as they could. A portal that allows users to sign-in to a profile, and customize that profile, could be a useful resource. A function should be implemented that allows users to build and edit workflows or training lists and save them for future use. This would drastically increase the teaching utility of the bioinformatics portal. Users need to be able to communicate with one another, form groups, and share training lists with those groups. An educator could compose a list of training materials and share that list with a group of students. This would also be a useful resource when presenting bioinformatics workshops. An application that can assess a user's intention, skill level, and needs would be a beneficial application as well. A user could sign-in, take an assessment, and have a generated list of training materials available to complete their desired intentions. For example, consider a user who wants to learn to perform a sequence alignment. Their assessment would indicate this need and generate a list of resources including

"Introduction to BioExtract Server," "ClustalW Usage," and "Phylogenic Tree Creation Tutorial" for example. Since the content is aimed at the user, a rating system for content based on user background would be advisable. Suggested categories of rating could be: high school student, post-secondary student, educator, and researcher.

Table 1 allows us to understand the BKTMS tools diversity. The diversity is based on Bioinformatics sub-domains and Bioinformatics applications. This shows that evaluation criteria for managing knowledge transmission are not yet a priority in BKTMS tools. Indeed, these BKTMS lack the factors necessary for the establishment of a complete pedagogical method. We can cite the factors as course level, monitoring and evaluation of the receiver (e.g. trainee, student), and many others. We propose a more formal study of these critical tools to contribute to the improvement of the next generations of BKTMS tools.

## 2. BKTMS Tools comparison
### a) Comparison principle

We use criteria provide in [10] because of its pluridisciplinarity and flexibility (Figure1). We provide a comparison of our review BKTMS (Table 1). That comparison involves 8 other LMS from EduTools and is described in [11].

Based on this information and our exploration of these tools, a grade table was proposed with KTMS in the rows and criteria in the columns (Table 2). Each criterion was weighted based on its importance related to the BKTMS. Based on the calculated scores, the user can select appropriate tools or use radar or histogram chart visualization for more details. The graphic visualization provides a snapshot of each KTMS position for each criterion (Figure 1 and Figure 2).

For simplification and compliance needs, grades for all criteria are brought to a discrete evaluation scale [1...8]. The numeric value of the evaluation may be changed, but with greatly improved ability to compare tools. The comparison is based on our context and our need. Each user can specify the priorities and degrees of importance to each criterion function in its own context. That is to say each criterion may have a score of 1 to 8, and a weight to express its importance in the given context.

### b) Score calculation and comparison result visualization

Scores were calculated using the simple expert's (decision maker's) additive utility function [12] (Eq.5)

$$f(X) = \sum_{i=1}^{m} p_i f_i(X) \quad \text{(Eq.5)}$$

where $f_i(X) = \{1,2,3,…,8\}$ is the rating grade of the criterion $i$ for each examined alternative BKTMS $X_j$, $m$ is the number of criteria, and $p_i$ is the weighted weight of a given criterion $i$ (weight i on the total weight).

### c) Grades, score calculation and score visualization.

Table 2 shows our BKTMS evaluation grades and scores. The first line contains the BKTMS name; lines 2 to 39 contain the KTMS grades for each criterion. The last line contains the score of each BKTMS weighting with the equation (Eq.5) formula. This table includes: grade for global criteria (white background) and detailed criteria (grey background), calculated score (black background) using equation (Eq.5), for global criteria (penultimate line) and detail criteria (last line). The first and second column contains the criterion name and its code, the third column contains the weight of the related criterion in our context. The remains columns contain the data related to each BKTMS. A high weight is put on pluridisciplinarity, collaboration, and networking criteria, given their predominance in bioinformatics.

Figure 2 shows radar chart visualization of KTMS global criteria scores (G) and detail criteria for each global criterion for more information: technical (T), pedagogy (P), interdisciplinary (I), communication (C), others (O) (related to Table 2).

## Conclusion

How can one choose an appropriate BKTMS? What BKTMS should be used for what goals? What training? What learning? What level? What is a bioinformatics program? We can continue the list indefinitely. Depending on one's involvement in the pluridisciplinary education world like bioinformatics, researcher can be looking for answers to some of these questions. This paper proposes some elements that can help in finding answers.

This study has primarily focused on providing a standardized critique of the existing BKTMS critics review. Even if the tools list was not exhaustive, it allowed us to present evidence of the strengths and failures of existing tools for bioinformatics knowledge transmission. Evaluation criteria were finally applied on the described BKTMS selected list.

Figure analysis has shown that the existing BKTMS has a relatively low score compared to the 8 selected LMS. And yet these BKTMS have high grades in multidisciplinary criteria. Through careful observation we noted that the grades for collaborativity, networking, and sharing criteria were low. Even strong weighting of the interdisciplinarity criteria could not compensate for this weakness.

This review and evaluation will be used to gain insight into the tools, systems, and capabilities that will be added to or excluded from a new proposed model for the next generation of BKTMS, involving web services and standards like SCORM, LOM, or IMS.

**Acknowledgments**: **Acknowledgments**: The authors would like to thank Dr. Doug Goodman and Jerry Prentice for their helpful corrections on this work.

**Funding**: This work was made possible by SD-INBRE Grant #P20RR016479-09 from the National Center for Research Resources (NCRR), a component of the National Institutes of Health (NIH). Its contents are solely the responsibility of the authors and do not necessarily represent the official views of NCRR or NIH. NSF Grant IOS-1126481 Integrating the BioExtract Server with the iPlant Collaborative.

**Terminologies:**
*LOs: Learning Objects*
*LORs: Learning Object Repository*
*CBT: Computer Based Training*
*LMS: Learning Management System*
*KTMS: Knowledge Transmission Management System*
*BKTMS: Bioinformatics Knowledge Transmission Management System*
*SCORM: Sharable Content Object Reference Model*
*VLEs: Virtual learning environments*
*CBTE: competence-based teaching education,*
*HBTE: humanistic-based teacher education,*
*IMS: instructional management system*
*LOM: learning object model*

**List of figures**

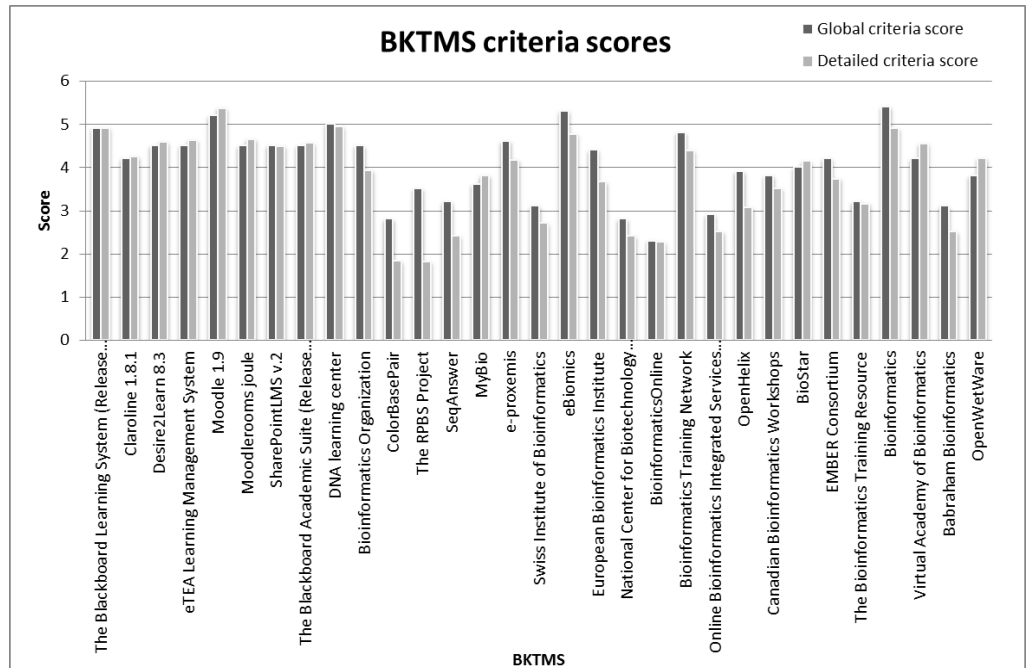

Figure 1: Score visualization histogram for BKTMS based on global evaluation criteria T,P,I,C,O (black), and detailed criteria (grey).

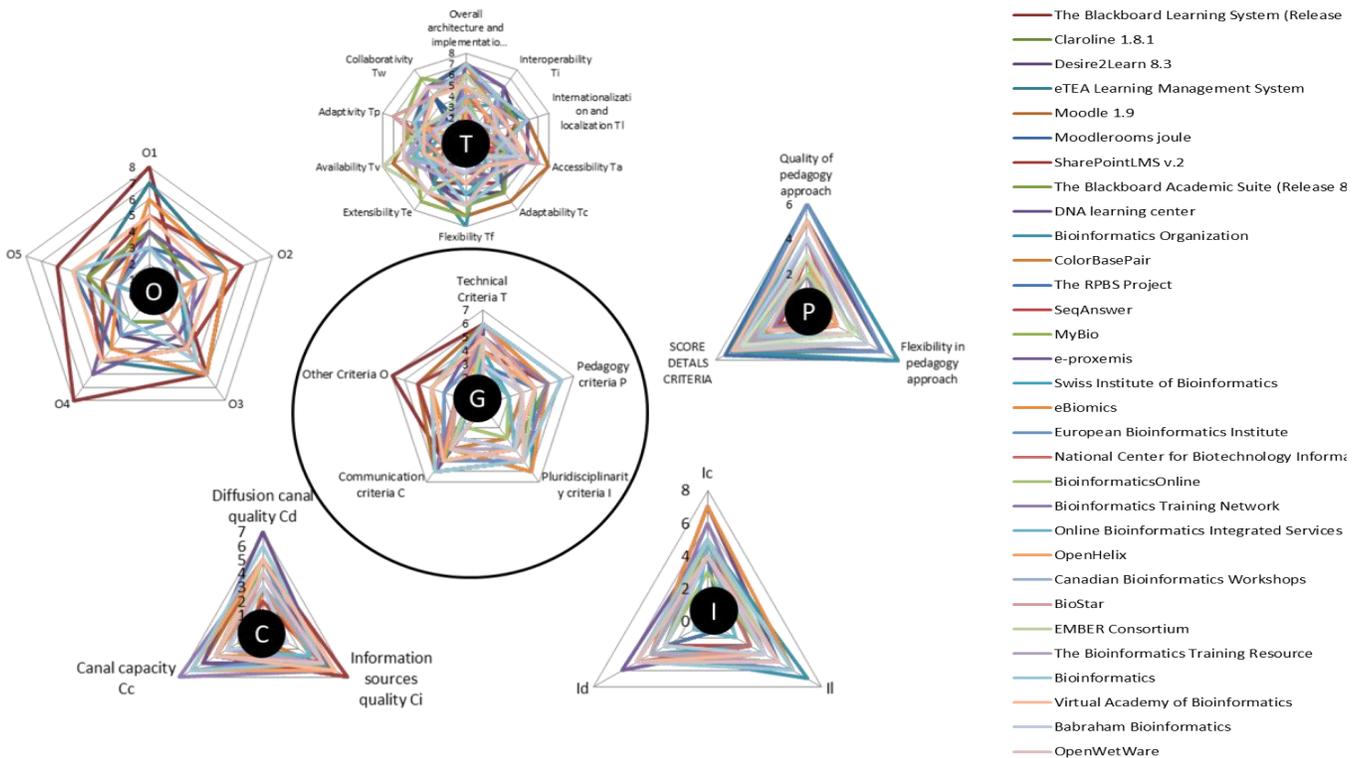

Figure 2: Score radar visualization of BKTMS based on global criteria (G). Zoom on detail criteria for each global criterion for more detail, technical (T), pedagogy (P), interdisciplinary (I), communication (C), others (O) (Table 4).

# List of tables:
Table 1: BKTMS overview

| Portal Name | Base Institution | Contact/author | Main Statement/Goals | URL | Comments | Approach | Login | Freely download, access materials | Updates clearly indicated | Reviewing/ranking material option | Materials searchable | Trainer/contact information | Info about training facility | Links to courses and events |
|---|---|---|---|---|---|---|---|---|---|---|---|---|---|---|
| Bioinformatics Organization | Bioinformatics Organization, Incorporated | J.W. Bizzaro | The Bioinformatics Organization develop computational resources to facilitate world-wide communications and collaborations between people of all educational and professional levels. It provide and promote open access to the materials and methods required for, and derived from, research, development and education. | http://www.bioinformatics.org/ | Basic usage is free, Professionals has a cost. Uses Wikipedia as a base site or reference location. | Unclear | Yes, not required. | Yes | Yes | No | Yes | Yes | No | Yes |
| Compbiology | | Jennifer Steinbachs | A computer biology related news site. | http://compbiology.org/ | Seems abandoned. | Unclear | Yes, not required. | NA | Yes | No | No | No | No | No |
| BioPlanet | ISCB | - | Collection of bioinformatics training programm in the world and job search | http://www.bioplanet.com/ | | Unclear | No | NA | No | No | No | No | No | Yes |
| ColorBasePair | - | - | Provide bioinformatics resources such as bioinformatics news, bioinformatics jobs, bioinformatics books, bioinformatics tutorials, bioinformatics training | http://www.colorbasepair.com/ | | Unclear | No | Yes | Yes | NA | No | No | No | No |
| MyBio | Wikia | | BIO international event planing tool (see and be seen, networking, explore sessions event) | http://mybio.wikia.com/wiki/Tutorials_in_bioinformatics | | collection list | Yes | Yes | Unclear | NA | Yes | No | No | No |
| e-proxemis | SIB | | Bioinformatics learning portal for proteomics | http://e-proxemis.expasy.org | | Try to fulfill resource requirements search by category of work being done. Similar for teaching tools. | Yes | NA | Unclear | No | No | No | No | No |
| Swiss Institute of Bioinformatics | | Peter Malama | collection of bioinformatics resources (slides, video) organize in programs, workshops, practices labs | | Display tools and offsite tools based on type of research [ex. sequence classification = PROSITE, Pfam, PRINTS, etc..). Has practice lab for tools such as BLAST. Connexion between events and training courses. | Foccus based organisation of resources | No | Yes | Yes | No | Yes | Yes | Yes | No |
| SEQanswers | USA and Chinese Universities | | SEQanswers provides a real-time knowledge-sharing resource to address this need, covering experimental and computational aspects of sequencing and sequence analysis | http://seqanswers.com/ | - | integration, collaborativity, question based, wiki | Yes | Yes | Yes | Yes | Yes | NA | NA | No |
| Galaxy | The Huck Institutes of the Life Sciences, The Institute for CyberScience at Penn State, and Emory University | | Galaxy is an open, web-based platform for *accessible*, *reproducible*, and *transparent* computational biomedical research. | http://wiki.galaxyproject.org/ | Workflow based tools and related learning resources | topic based videos, wiki | Yes | Yes | Yes | No | Yes | NA | NA | NA |
| eBiomics | NBIC / Wageningen / ULB / SDC | many persons | Devoted to workbench basedn training in bioinformatics. It is composed of several interconnected sections that can be accessed through different interactive activities. The purpose of eBiomics is to familiarise users with bioinformatics analysis flows in diverse -omics applications. | http://ebiomics.sdcinfo.com/ | For each database or tool, they have "how to use this" sections, and a link to the tool. | Try to fulfill resource requirements search by category of work being done. Similar for teaching tools. | Yes | Yes | Yes | No | Yes | Yes | No | Yes |
| Bioinformatics Resource Portal | | | providing resources | http://bioinformaticstools.webs.com/index.htm | Many tools and databases are provided. | | No | Yes | No | No | No | No | No | No |
| European Bioinformatics Institute | European Molecular Biology Laboratory | Janet Thornton | Provide advanced bioinformatics training to scientists at all levels, from PhD students to independent investigators. | http://www.ebi.ac.uk/ | | Primary database. Find resources and training tools by research category. | | | | | | | | |
| GATK forum | | | GATK is an industrial-strength infrastructure and engine that handle data access, conversion and traversal, as well as high-performance computing features | http://gatkforums.broadinstitute.org/ | | Primary database. Find resources and training tools by research category. | Yes | Yes | Yes | No | Yes | No | No | No |
| National Center for Biotechnology Information | NCBI | board of members | collection of learning resources for tools and databases use in NCBI portal | http://www.ncbi.nlm.nih.gov/guide/training-tutorials/ | | Primary database. Find resources and training tools by research category. | No | Yes | Unclear | No | No | No | No | No |
| BioinformaticsOnline | Bioinformatics Skill Development Programme | Jitendra Narayan | BioinformaticsOnline(BOL) is a bioinformatics education portal for the students of Bioinformatics and Biotechnology and specifically designed to help research scientist, working on various project. The aims at bringing together research scientists interested in Bioinformatics and allied fields and to support develop and spread Bioinformatics in academic, technologic and industrial environment in India and abroad. | http://www.bioinformaticsonline.com/ | | Separate by tools, databases, career, training, companies, etc... (not based off of discipline or anything) | Yes | Yes | No | No | Yes | No | No | No |
| Bioinformatics Training Network | EMBL-EBI based maintainers | EMBL-EBI | The BTN is a community-based project aiming to provide a centralised facility to share materials, to list training events (including course contents and trainers), and to share and discuss training experiences. You are welcome to browse the site, and we encourage you to join us to share (license) information and materials – please register. | http://www.biotnet.org/ | Devoted to the teachers/professors/trainers | Provide teaching resources based off of research category (proteomics, nutrigenomics, etc.), also list available workshops, and list training facilities. | Yes | Yes | Yes | No | Yes | Yes | Yes | Yes |
| Online Bioinformatics Integrated Services (iBIS) | University of Miami and OpenHelix | University of Miami | iBIS is UM's online Bioinformatics Integrated Services portal. iBIS was developed particularly for biologists and medical researchers- it is a user-friendly, (fairly) comprehensive, customizable portal which includes education materials to guide you through your bioinformatics experience. The idea is that your saved workspace in iBIS contains links to all your frequently used bioinformatics databases, tools, or UM services. | http://bio.ccs.miami.edu/ibis/UMIBIS.jsp | Student resources available through OpenHelix. | Use of widgets to allow easy access of tools, databases, etc. Allows users to "save workspace" or list of tools present for certain research or training resources | Yes | No) | No | No | No | No | No | No |
| Bioportal RENCI | | Stan Ahalt | We bring the latest cyber tools and technologies to bear on pressing problems. We work with scientists who study critical issues from climate change to the causes of cancer. We form research teams that involve faculty members at universities across North Carolina and the U.S. and that are positioned to bring research projects to North Carolina. We partner with North Carolina government agencies so they are able to better serve the state's citizens. | http://www.renci.org/ | | More a private group advancing computational technology. | No | NA | Yes | NA | Yes | Yes | NA | Yes |
| Bioinformatics Institute of India (BII) | - | - | Teaching for money | http://www.bii.in/ | | Pay to train or learn bioinformatics | Yes | No | Yes | Yes | Yes | Yes | Yes | Yes |
| Oslo Bioportal | University of Oslo | - | The Bioportal at University of Oslo is a web-based portal for phylogenomic analysis, population genetics and high-throughput sequence analysis. The main advantage of Bioportal lies in an access to a parallel computational resource that enables demanding computations. Therefore, this resource is designed for large, time consuming computations rather than for an interactive use. | http://www.bioportal.uio.no/ | | Have some tools and instructions for use for phylogenomic analysis. Based off of tool name. | Yes | Yes | Yes | Yes | Yes | Yes | Yes | Yes |
| OpenHelix | own company | Mary E. Mangan | OpenHelix empowers researchers by providing a search portal to find the most relevant genomics resource and training on those resources; distributing extensive and effective tutorials and training materials on the most powerful and popular genomics resources; contracting with resource providers to provide comprehensive, long-term training and outreach programs. | http://www.openhelix.com/ | Must pay for training resources. | Based off of ABC order of tool or database names. Have some free resources. | No | Partially | Yes | No | Yes | Yes | No | No |
| Canadian Bioinformatics Workshops | bioinformatics.ca and supporters | Michelle Brazas | Provide workshops | http://www.bioinformatics.ca/ | Have won many education awards, Might have to pay to attend workshops | Host workshops around Canada and British Columbia on different bioinformatics topics. Such as: cancer genomics, metabolomics, thorough sequencing, microarray data analysis. | No | NA | No | No | No | No | No | No |
| eusol | European Commission | Dr. R.M. Klein Lankhorst | The distributed bioinformatics platform aims at making state-of-the-art bioinformatics data and analyses available to all EU-SOL partners, which will be offered training in utilizing this resource for their research. | http://www.eu-sol.net/science/bioinformatics/portal-overview | Tomatoes information based. | Provide tools, training of those tools, and databases related to bioinformatics on tomatoes. | Yes | Yes | Yes | Yes | Yes | Yes | Yes | Yes |
| BioStar | Unclear | Parnell Lindenbaum | is site's focus is bioinformatics, computational genomics and biological data analysis. We welcome posts that are detailed and specific, written clearly and simply, of interest to at least one other person somewhere | http://www.biostars.org/show/questions/ | Question answers rating forum | More of a question forum for bioinformatics. | Yes | Yes | Yes | Yes | Yes | NA | NA | NA |
| ExPASy | SIB | SIB | ExPASy is the SIB Bioinformatics Resource Portal which provides access to scientific databases and software tools (i.e., resources) in different areas of life sciences including proteomics, genomics, phylogeny, systems biology, population genetics, transcriptomics etc. On this portal you find training resources from many different SIB groups as well as external institutions. | http://expasy.org/ | | | No | Yes | Yes | No | Yes | Yes | No | Yes |
| EMBER Consortium | EMBER | | online practical training resources designed to introduce a range of bioinformatics services, databases and software available on the Web | http://www.ember.man.ac.uk | | Course, Question, QUIZZ | Yes | Yes | NA | No | Yes | NA | NA | NA |
| The Bioinformatics Training Resource | BTR | | BTR is an organized collection of links to online tutorials, online courses, essays, book chapters, course syllabi, glossaries, bibliographies of key papers, etc. In short everything that interested scientists need in order to train themselves in the emerging discipline of bioinformatics. | http://www.med.nyu.edu/rcr/rcr/btr/ | | | No | Yes | No | No | Yes | No | No | Yes |
| Bioinformatics | | | professional development courses for continuing scientific education. | http://www.bioinformatics.org/edu/ | use Moodle LMS tools | use Moodle site based tools | Yes | No | Yes | No | Yes | No | No | Yes |
| Virtual Academy of Bioinformatics | BioInfoBank | | Bioinformatics master degree programs collection of courses (and lectures inside courses) built around a subject. | http://lib.bioinfo.pl/programs/view/1 | | Formal academic course approach | Yes | Yes | Yes | No | Yes | Yes | Yes | Yes |
| Babraham Bioinformatics | Babraham Institute | | Collection of courses related to 30 bioinformatics research teams collaboration | http://www.bioinformatics.babraham.ac.uk/training.html | | Research topic focus tutorial | No | yes | No | No | Yes | Yes | Yes | Yes |
| OpenWetWare | Collaborative effort | | OpenWetWare is an effort to promote the sharing of information, know-how, and wisdom among researchers and groups who are working in biology & biological engineering | http://openwetware.org/wiki/Wikiomics | | Wiki and collaborative | Yes | Yes | Yes | Yes | Yes | Yes | Yes | No |
| DNA Learning Center | | | The DNA Learning Center (DNALC) is the world's first science center devoted entirely to genetics education and is an operating unit of Cold Spring Harbor Laboratory, an important center for molecular genetics research. | http://www.dnalc.org/about/ | | cognitive teaching, 3D course | Yes | No | Yes | Yes | Yes | Yes | Yes | NA |

Table 2: BKTMS table for evaluation grades and scores calculation.

| Criteria | | | The Blackboard Learning System (Release 7) - Enterprise License | Claroline 1.8.1 | Desire2Learn 8.3 | eTEA Learning Management System | Moodle 1.9 | Moodlerooms joule | SharePointLMS v.2 | The Blackboard Academic Suite (Release 8.0) | DNA learning center | Bioinformatics Organization | ColorBasePair | The RPBS Project | SeqAnswer | MyBio | e-proxemis | Swiss Institute of Bioinformatics | eBiomics | European Bioinformatics Institute | National Center for Biotechnology Information | BioinformaticsOnline | Bioinformatics Training Network | Online Bioinformatics Integrated Services (iBIS) | OpenHelix | Canadian Bioinformatics Workshops | BioStar | EMBER Consortium | The Bioinformatics Training Resource | Virtual Academy of Bioinformatics | Babraham Bioinformatics | OpenWetWare |
|---|---|---|---|---|---|---|---|---|---|---|---|---|---|---|---|---|---|---|---|---|---|---|---|---|---|---|---|---|---|---|---|---|
| Overall architecture and implementation | To | 1 | 7 | 6 | 7 | 7 | 7 | 7 | 5 | 7 | 6 | 6 | 3 | 2 | 4 | 5 | 7 | 3 | 7 | 5 | 4 | 4 | 6 | 4 | 5 | 4 | 6 | 6 | 3 | 7 | 5 | 3 | 5 |
| Interoperability | Ti | 1 | 5 | 5 | 6 | 4 | 5 | 5 | 4 | 5 | 5 | 5 | 0 | 1 | 1 | 4 | 6 | 4 | 4 | 3 | 3 | 2 | 5 | 5 | 3 | 5 | 4 | 4 | 2 | 5 | 4 | 2 | 4 |
| Internationalization and localization | Tl | 1 | 6 | 5 | 5 | 6 | 6 | 6 | 5 | 5 | 6 | 2 | 0 | 0 | 3 | 3 | 4 | 4 | 2 | 2 | 4 | 3 | 3 | 3 | 5 | 4 | 2 | 6 | 4 | 3 | 3 |
| Accessibility | Ta | 1 | 3 | 4 | 4 | 3 | 8 | 3 | 3 | 4 | 6 | 7 | 4 | 5 | 4 | 5 | 3 | 2 | 5 | 5 | 3 | 4 | 5 | 4 | 2 | 6 | 7 | 6 | 5 | 3 | 6 | 5 | 6 |
| Adaptability | Tc | 1 | 6 | 4 | 6 | 6 | 7 | 5 | 5 | 6 | 5 | 2 | 0 | 1 | 1 | 3 | 4 | 3 | 3 | 3 | 1 | 2 | 4 | 4 | 2 | 3 | 2 | 3 | 3 | 5 | 3 | 2 | 4 |
| Flexibility | Tf | 1 | 5 | 5 | 5 | 6 | 7 | 5 | 4 | 6 | 6 | 8 | 0 | 0 | 1 | 7 | 3 | 3 | 3 | 2 | 1 | 2 | 4 | 5 | 1 | 4 | 3 | 3 | 3 | 6 | 4 | 3 | 6 |
| Extensibility | Te | 1 | 5 | 5 | 5 | 5 | 6 | 5 | 5 | 5 | 5 | 5 | 0 | 0 | 4 | 7 | 4 | 4 | 4 | 3 | 4 | 2 | 4 | 3 | 2 | 3 | 3 | 6 | 4 | 4 | 3 | 3 | 6 |
| Availability | Tv | 1 | 3 | 4 | 3 | 3 | 7 | 4 | 4 | 4 | 4 | 4 | 2 | 4 | 4 | 5 | 4 | 4 | 4 | 2 | 2 | 6 | 6 | 5 | 2 | 6 | 3 | 8 | 6 | 4 | 4 | 4 | 6 |
| Adaptivity | Tp | 1 | 5 | 5 | 5 | 5 | 5 | 6 | 5 | 5 | 4 | 2 | 2 | 4 | 6 | 3 | 3 | 2 | 4 | 5 | 1 | 3 | 4 | 3 | 4 | 7 | 4 | 5 | 4 | 3 | 6 |
| Collaborativity | Tw | 5 | 6 | 5 | 6 | 5 | 6 | 6 | 5 | 5 | 3 | 1 | 1 | 2 | 7 | 2 | 2 | 5 | 2 | 0 | 2 | 6 | 3 | 0 | 3 | 5 | 3 | 3 | 5 | 5 | 2 | 6 |
| LO interoperability | Li | 1 | 5 | 5 | 5 | 6 | 6 | 5 | 4 | 5 | 4 | 2 | 0 | 0 | 2 | 3 | 3 | 2 | 5 | 3 | 1 | 1 | 3 | 2 | 3 | 5 | 6 | 3 | 2 | 6 | 3 | 3 | 6 |
| LO contextualization | Lc | 1 | 6 | 6 | 6 | 5 | 7 | 4 | 4 | 6 | 4 | 1 | 2 | 2 | 1 | 6 | 8 | 5 | 4 | 2 | 4 | 3 | 5 | 4 | 4 | 3 | 5 | 2 | 4 | 4 | 3 | 6 |
| LO Diversity | Ll | 1 | 5 | 5 | 5 | 4 | 4 | 5 | 5 | 5 | 5 | 3 | 3 | 3 | 4 | 2 | 3 | 3 | 2 | 3 | 3 | 1 | 2 | 4 | 5 | 3 | 3 | 6 | 6 | 1 | 6 |
| LO Accessibility | La | 1 | 3 | 4 | 4 | 4 | 6 | 4 | 3 | 4 | 4 | 5 | 3 | 3 | 5 | 3 | 1 | 5 | 5 | 4 | 4 | 5 | 3 | 2 | 6 | 6 | 5 | 5 | 3 | 4 | 2 | 6 |
| LO architecture | Lp | 1 | 6 | 5 | 5 | 5 | 6 | 6 | 4 | 5 | 5 | 1 | 1 | 1 | 3 | 6 | 3 | 4 | 4 | 2 | 2 | 4 | 2 | 5 | 5 | 4 | 6 | 3 | 5 | 4 | 3 | 4 |
| LO Design and usability | Lu | 1 | 6 | 4 | 5 | 5 | 5 | 5 | 5 | 5 | 4 | 1 | 2 | 4 | 3 | 7 | 3 | 4 | 6 | 3 | 2 | 4 | 2 | 4 | 4 | 5 | 4 | 7 | 5 | 2 | 4 |
| LO Interactivity | Ls | 1 | 4 | 5 | 5 | 5 | 6 | 5 | 5 | 5 | 6 | 2 | 1 | 1 | 2 | 2 | 3 | 2 | 4 | 6 | 4 | 3 | 4 | 1 | 3 | 3 | 7 | 4 | 3 | 6 | 5 | 3 | 3 |
| LO Verification ability | Lv | 1 | 5 | 5 | 5 | 5 | 5 | 4 | 5 | 4 | 3 | 1 | 1 | 0 | 2 | 5 | 3 | 6 | 6 | 5 | 3 | 6 | 4 | 5 | 4 | 5 | 5 | 4 | 5 | 6 | 4 | 6 |
| LO tagging ability | Lt | 1 | 6 | 5 | 6 | 6 | 5 | 5 | 5 | 5 | 4 | 0 | 0 | 0 | 1 | 5 | 1 | 2 | 1 | 6 | 3 | 2 | 2 | 6 | 3 | 2 | 6 | 2 | 6 | 4 | 1 | 4 |
| LO Retrievability | Lr | 1 | 5 | 5 | 5 | 5 | 6 | 5 | 5 | 5 | 5 | 2 | 4 | 4 | 4 | 3 | 6 | 3 | 4 | 3 | 3 | 3 | 4 | 1 | 4 | 4 | 4 | 3 | 2 | 4 | 5 | 1 | 4 |
| LO Discipline dependence | Ld | 4 | 4 | 4 | 4 | 4 | 4 | 4 | 4 | 4 | 4 | 7 | 6 | 3 | 3 | 3 | 4 | 5 | 2 | 4 | 3 | 2 | 3 | 1 | 3 | 2 | 4 | 3 | 3 | 5 | 6 | 1 | 4 |
| **Technical Criteria** | **T** | 6 | 5 | 5 | 5 | 6 | 4 | 5 | 6 | 3 | 3 | 3 | 5 | 6 | 4 | 6 | 3 | 3 | 5 | 3 | 4 | 5 | 6 | 5 | 3 | 6 | 5 | 3 | 5 |
| Quality of pedagogy approach | Pa | 1 | 5 | 5 | 5 | 5 | 5 | 5 | 4 | 5 | 4 | 3 | 3 | 3 | 3 | 4 | 2 | 6 | 4 | 3 | 2 | 5 | 3 | 4 | 3 | 2 | 4 | 3 | 5 | 4 | 3 |
| Flexibility in pedagogy approach | Pf | 6 | 5 | 5 | 5 | 5 | 6 | 5 | 5 | 5 | 6 | 3 | 1 | 1 | 2 | 3 | 3 | 2 | 5 | 1 | 1 | 4 | 2 | 3 | 3 | 4 | 3 | 3 | 4 | 4 | 3 | 3 |
| **Pedagogy criteria** | **P** | 3 | 5 | 5 | 5 | 5 | 5 | 5 | 5 | 6 | 4 | 3 | 4 | 3 | 3 | 4 | 2 | 5 | 2 | 2 | 5 | 2 | 3 | 4 | 4 | 3 | 6 | 4 | 3 | 3 |
| Pluridisciplinarity cognitive map creation | Ic | 5 | 4 | 5 | 4 | 4 | 4 | 4 | 4 | 5 | 7 | 5 | 4 | 4 | 4 | 5 | 4 | 7 | 5 | 4 | 3 | 6 | 5 | 4 | 4 | 5 | 5 | 4 | 4 | 4 |
| Pluridisciplinarity learning language | Il | 4 | 4 | 4 | 4 | 4 | 5 | 4 | 4 | 4 | 5 | 7 | 1 | 1 | 3 | 4 | 3 | 3 | 6 | 6 | 3 | 4 | 4 | 2 | 4 | 5 | 4 | 5 | 5 | 6 | 4 | 4 | 5 |
| Dominant idea definition | Id | 4 | 3 | 3 | 3 | 5 | 3 | 3 | 3 | 4 | 3 | 3 | 3 | 6 | 4 | 4 | 4 | 5 | 4 | 1 | 5 | 5 | 4 | 4 | 4 | 5 | 5 | 3 | 5 |
| **Pluridisciplinarity criteria** | **I** | 4 | 4 | 4 | 4 | 4 | 5 | 4 | 4 | 4 | 5 | 6 | 3 | 4 | 4 | 4 | 5 | 4 | 6 | 5 | 4 | 3 | 5 | 4 | 4 | 5 | 4 | 5 | 4 | 5 | 4 | 4 | 5 |
| Diffusion canal quality | Cd | 1 | 5 | 5 | 5 | 5 | 5 | 5 | 5 | 5 | 7 | 4 | 3 | 2 | 2 | 5 | 4 | 4 | 3 | 5 | 3 | 1 | 5 | 4 | 5 | 4 | 4 | 3 | 3 | 6 | 5 | 3 | 4 |
| Information sources quality | Ci | 1 | 7 | 6 | 6 | 6 | 6 | 6 | 7 | 6 | 5 | 5 | 6 | 5 | 6 | 6 | 5 | 4 | 4 | 2 | 2 | 6 | 4 | 6 | 5 | 5 | 5 | 5 | 5 | 6 | 2 | 5 |
| Canal capacity | Cc | 1 | 4 | 4 | 4 | 4 | 5 | 4 | 4 | 4 | 5 | 3 | 1 | 2 | 2 | 3 | 4 | 3 | 4 | 6 | 3 | 2 | 7 | 3 | 6 | 3 | 3 | 3 | 3 | 6 | 4 | 3 | 4 |
| **Communication criteria** | **C** | 1 | 5 | 5 | 5 | 5 | 6 | 5 | 5 | 5 | 6 | 4 | 4 | 4 | 4 | 5 | 5 | 4 | 4 | 5 | 3 | 2 | 6 | 3 | 5 | 4 | 4 | 4 | 4 | 6 | 5 | 3 | 4 |
| Online Gradebook | O1 | 1 | 8 | 0 | 4 | 7 | 5 | 3 | 5 | 4 | 0 | 0 | 0 | 0 | 0 | 0 | 4 | 0 | 6 | 3 | 0 | 0 | 0 | 0 | 0 | 0 | 0 | 0 | 0 | 3 | 5 | 0 | 0 |
| Student Tracking | O2 | 1 | 2 | 1 | 3 | 5 | 6 | 5 | 6 | 3 | 0 | 0 | 2 | 0 | 0 | 0 | 0 | 4 | 1 | 5 | 3 | 0 | 0 | 3 | 0 | 3 | 0 | 0 | 2 | 0 | 2 | 4 | 0 | 0 |
| Real-time Chat | O3 | 1 | 6 | 1 | 4 | 6 | 4 | 6 | 4 | 2 | 0 | 4 | 0 | 0 | 0 | 3 | 3 | 0 | 1 | 6 | 2 | 0 | 0 | 4 | 0 | 2 | 0 | 6 | 0 | 0 | 5 | 4 | 0 | 0 |
| Automated Testing Management | O4 | 1 | 8 | 0 | 3 | 5 | 5 | 4 | 5 | 2 | 0 | 0 | 0 | 0 | 0 | 0 | 6 | 1 | 4 | 3 | 0 | 0 | 0 | 0 | 0 | 0 | 0 | 0 | 0 | 2 | 5 | 0 | 0 |
| Self-assessment | O5 | 1 | 6 | 1 | 3 | 4 | 4 | 3 | 4 | 4 | 0 | 0 | 0 | 0 | 0 | 0 | 2 | 2 | 1 | 2 | 2 | 0 | 2 | 2 | 0 | 0 | 0 | 0 | 0 | 5 | 5 | 0 | 0 |
| **Other Criteria** | **O** | 1 | 7 | 1 | 4 | 4 | 5 | 4 | 5 | 4 | 0 | 1 | 0 | 0 | 0 | 1 | 3 | 1 | 4 | 3 | 0 | 0 | 2 | 1 | 2 | 0 | 2 | 1 | 0 | 4 | 4 | 0 | 0 |
| **SCORE GLOBAL CRITERIA** | | 10 | 4.9 | 4.2 | 4.5 | 4.5 | 5.2 | 4.5 | 4.5 | 4.5 | 5 | 4.5 | 2.8 | 3.5 | 3.2 | 3.6 | 4.6 | 3.1 | 5.3 | 4.4 | 2.8 | 2.3 | 4.8 | 2.9 | 3.9 | 3.8 | 4 | 4.2 | 3.2 | 5.4 | 4.2 | 3.1 | 3.8 |
| **SCORE DETAILS CRITERIA** | | 57 | 4.9 | 4.2 | 4.6 | 4.6 | 5.4 | 4.6 | 4.5 | 4.6 | 4.9 | 3.9 | 1.8 | 1.8 | 2.4 | 3.8 | 4.2 | 2.7 | 4.8 | 3.7 | 2.4 | 2.3 | 4.4 | 2.5 | 3.1 | 3.5 | 4.1 | 3.7 | 3.2 | 4.9 | 4.5 | 2.5 | 4.2 |